\documentclass[12pt,a4paper]{amsart}
\usepackage[french]{babel}
\usepackage[latin1]{inputenc}
\usepackage[T1]{fontenc}
\usepackage{mathrsfs,dsfont}
\usepackage{amssymb,amsmath}
\usepackage{url}

\def\pacs#1{\textbf{P.A.C.S.:}\ #1}

\begin{document}
\title{Mécanique ondulatoire et C-équivalence}
\author{Satyanad Kichenassamy}
\address{Laboratoire de Math\'ematiques, Universit\'e de Reims Champagne-Ardenne, Moulin de la Housse, B.P. 1039, F-51687 Reims Cedex 2, France.}

\email{satyanad.kichenassamy@univ-reims.fr}%

\date{6 juin 2019 (June~6, 2019)}

\vskip 1em

\hfill%
\emph{Annales de la Fondation Louis de Broglie} {\bf 45}(1)  (2015), 99--111.

\vskip 0.5em

\hfill
\url{ https://fondationlouisdebroglie.org/AFLB-451/aflb451m922.htm}

\vskip 2em

\large
\maketitle

\markboth{Mécanique ondulatoire et C-équivalence}{Mécanique ondulatoire et C-équivalence}

\vskip 1.5cm
\begin{abstract}
On montre que l'onde de Broglie, telle qu'elle est introduite dans sa Thèse, n'est pas une onde sur un espace préexistant, mais d\'etermine elle-même un syst\`eme propre dans lequel la Relativit\'e restreinte est localement valable. Ce syst\`eme est un objet physique qui définit ses propres unités de temps et d'espace. La C-\'equivalence fournit un cadre naturel pour sa description mathématique et son interprétation physique.

{\it 
ABSTRACT.  We establish that de Broglie's wave, as it is introduced in his Thesis, is not a wave on a given space, but on the contrary itself determines a system in which Special Relativity is locally valid. This local system is a physical object, that defines its own units of length and time. C-equivalence provides a natural framework both for its mathematical description and its physical interpretation.}
\end{abstract}


\vskip 2em

\section{Introduction}

Si les prédictions de la Mécanique ondulatoire ont reçu les confirmations que l'on sait, la vision de Louis de Broglie n'a jamais reçu la formulation mathématique cohérente qu'il n'avait cessé d'appeler de ses v\oe ux. On montre, par une lecture attentive de passages de sa Thèse \cite{these-LdB}, que l'onde de phase d'un \og morceau d'énergie\fg\ est pour lui une réalité physique définissant son système propre, que nous interprétons comme un pseudo-système d'inertie au sens de la C-équivalence \cite{K-ihp-ceq,c-equiv}. Autrement dit, c'est pour nous l'objet physique qui définit son espace et son temps. [p.~100] La considération de l'ensemble de ces pseudo-systèmes suggère une critique de la notion mathématique de variété telle qu'elle fut développée après de Broglie.

Il est à peine nécessaire de rappeler que nous ne disposons toujours pas d'une formulation mathématique cohérente de la Mécanique ondulatoire, comme l'a récemment souligné avec lucidité M. C. Cohen-Tannoudji  :
\begin{quote}
\og Une pratique quotidienne de cette discipline permet au physicien d'acquérir une certaine maîtrise dans le maniement du formalisme quantique, une certaine intuition des erreurs qu'il ne faut pas commettre, de la démarche qu'il faut suivre pour arriver au résultat correct. Cependant, lequel d'entre nous n'a pas ressenti, au moins une fois dans son parcours scientifique, un certain trouble concernant les fondements de la mécanique quantique, l'impression qu'une formulation convaincante et satisfaisante de cette théorie restait encore à élaborer ? \fg
\end{quote}
L'ouvrage qu'il préfaçait en ces termes considérait bien les développements issus de la Thèse de Louis de Broglie, mais seulement au travers de l'interprétation de Bohm. M. Laloë y décrivait les limitations de cette dernière en ces termes  : si elle \og réussit brillamment à éliminer le rôle de l'observateur au cours des mesures \fg \cite[p.~389]{cct-laloe} en introduisant deux \og niveaux de réalité \fg,
\begin{quote}
\og [l]e fait même que les deux niveaux de réalité soient distingués par les effets possibles des manipulations par les expérimentateurs montre que l'observateur, qu'on avait espéré faire disparaître de la théorie, s'insère à nouveau dans la description des phénomènes physiques. De plus on remarque que la tension [...] qui existe entre relativité et mécanique quantique standard trouve son image miroir dans la théorie dBB [de Broglie-Bohm] : c'est bien la relativité qui contraint à postuler une impossibilité fondamentale de manipuler les variables bohmiennes [...]\fg.
\end{quote}
Louis de Broglie, on le sait, a très tôt souligné la place centrale de la Relativité restreinte dans sa pensée, ainsi que le lien entre le protocole de mesure et les résultats qu'ils fournissent. Il a par ailleurs plaidé pour l'exclusion de la subjectivité de la mesure physique \cite{temps-nietzsche}. Pour autant, sa théorie ne vise pas à éliminer ce que nous appelons ici un observateur c'est-à-dire, comme en Relativité restreinte, un ensemble de protocoles [p.~101] permettant, potentiellement ou effectivement, d'assigner dans un premier temps des coordonnées à une classe d'événements, et dans un second temps de déterminer les relations de celles-ci avec les grandeurs physiques \cite{sk-cours}.

Nous montrons que la C-équivalence, bien qu'introduite pour obtenir une interprétation physique cohérente de la Relativité générale, théorie classique par essence \cite{anderson}, reste pertinente dans le domaine microscopique, pourvu que l'on remette en question la notion de variété telle que l'a définie Whitney en 1936 \cite{whitney}. \'Elie Cartan, dans la deuxième édition (1946) de son cours de 1925-26, ne semblait pas convaincu par celle-ci : \og La notion générale de variété est assez difficile à définir avec précision \fg, disait-il, avec un renvoi à divers traités et mémoires entre 1893 et 1940 \cite{cartan-varietes}, où celui de Whitney ne figure pas. La conception de la variété des observateurs, telle que la concevait Kichenassamy, et que nous prolongeons, fut nourrie entre autres des travaux de 1923 d'\'Elie Cartan \cite{cartan-aens}, dans lesquels il appelait de ses v\oe ux le développement d'une mathématique des observateurs.

Si l'on peut raisonnablement imaginer un observateur macroscopique construisant des coordonnées autour de lui, du moins dans un \og voisinage limité \fg, il n'en est pas de même dans le domaine microscopique : comment un méson déterminerait-il un espace de Minkowski local ? Et pourtant, la Relativité restreinte montre bien que sa durée de vie propre diffère en général de celle mesurée dans le système du laboratoire. Comment alors définir physiquement le système propre du méson sans référence au système du laboratoire ?

\section{Le système propre dans la Thèse de Louis de Broglie}

La Thèse de Broglie montre qu'il n'est pas nécessaire de faire intervenir un sujet conscient pour envisager un observateur local. Chacun se souvient de l'image (\cite{these-LdB}, Ch.~I, \S I, p.~23 sqq.) d'un \og plateau circulaire horizontal de très grand rayon\fg\ :
\begin{quote}
	\og à ce plateau, sont suspendus des systèmes identiques formés d'un ressort spiral auquel est accroché un poids. [...] Tous les systèmes ressorts-poids étant identiques ont tous même période ; faisons-les osciller avec la même amplitude et la même phase. La surface passant par les centres de gravité de tous les poids sera un plan qui montera et descendra d'un mouvement alternatif. L'ensemble ainsi obtenu présente une très grossière [p.~102] analogie avec le morceau isolé d'énergie tel que nous le concevons. \fg\
\end{quote}
Il fut conduit à cette conception en confrontant la théorie de Bohr et la Relativité restreinte : si un système comprend une famille caractéristique d'oscillateurs, il dispose d'un jeu d'unités de temps privilégiées. Mais si cet objet n'est pas strictement ponctuel, il faut bien postuler un système physique et non fictif dans lequel ces oscillateurs soient au repos relatif, et mis à l'heure, de manière à présenter entre eux des relations de phase bien définies. Pour tout autre observateur de la Relativité Restreinte, ce système d'oscillateurs sera représenté par une onde à cause des propriétés de la transformation de Lorentz. D'un point de vue mathématique, cela revient à dire que la phase est un scalaire pour les transformations de Lorentz : une expression telle que $\exp(i\omega t)$ dans un système deviendra, dans tel autre, $\exp(i\omega\gamma(t'-vx'/c^2))$, avec des notations évidentes. Nous aboutissons ainsi à un \textbf{premier point} :
\begin{quote}
Pour Louis de Broglie, postuler l'existence de l'onde de phase, c'est affirmer que le système propre d'un \og morceau d'énergie \fg, même microscopique, est un objet physique, étendu, et qu'il est, en l'absence d'accélérations, un système d'inertie de la Relativité restreinte. C'est un système distribué d'oscillateurs réels et non d'individus fictifs.
\end{quote}
Ce point est confirmé et complété par le passage suivant, où il envisage l'effet d'accélérations éventuelles (Ch.~II, \S VI, p.~46) :
\begin{quote}
\og Souvenons-nous maintenant des résultats obtenus au chapitre précédent dans le cas du mouvement uniforme. Nous avions alors été amenés à considérer l'onde de phase comme due aux intersections par l'espace actuel de l'observateur fixe des espaces passés, présents et futurs de l'observateur entraîné. Nous pourrions être tentés ici encore de retrouver la valeur donnée ci-dessus de $V$ [la vitesse de phase $c^2/v$] en étudiant les \og phases \fg\ successives du mobile et en précisant le déplacement pour l'observateur fixe des sections par son espace des états équiphases. Par malheur, on se heurte ici à de très grosses difficultés. La Relativité ne nous apprend pas actuellement comment un observateur entraîné par un mouvement non uniforme découpe à chaque instant son espace dans l'espace-temps ; il ne semble pas qu'il y ait beaucoup de raisons pour que cette section soit plane comme dans [p.~103] le mouvement uniforme. Mais si cette difficulté était résolue, nous serions encore dans l'embarras. En effet, un mobile en mouvement uniforme doit être décrit de la même façon par l'observateur qui lui est lié, quelle que soit la vitesse du mouvement uniforme, par rapport à des axes de référence ; cela résulte du principe que des axes galiléens possédant les uns par rapport aux autres des mouvements de translation uniforme sont équivalents. Si donc notre mobile est entouré, pour un observateur lié, d'un phénomène périodique ayant partout même phase, il doit en être de même pour toutes les vitesses du mouvement uniforme et c'est ce qui justifie notre méthode du chapitre premier. Mais si le mouvement n'est pas uniforme, la description du mobile faite par l'observateur lié peut n'être plus la même et nous ne savons plus du tout comment il va définir le phénomène périodique et s'il lui attribuera même phase en tout point de l'espace. [\dots] Nous ne pouvons aborder ce difficile problème. \fg
\end{quote}
Lorsqu'il s'agit de mouvements non uniformes, le système inertiel instantanément lié ne coïncide donc pas avec le système entraîné, ni en général avec les systèmes osculateurs ou \og sur-osculateurs\fg\ que l'on pourrait envisager.
Il faut donc associer à chaque observateur et à chaque phase de son évolution un espace-temps \emph{a priori} différent, dont la nature mathématique reste à définir, ainsi qu'un protocole pour comparer les mesures d'observateurs voisins. C'est ce qu'a accompli la C-équivalence pour les observateurs communiquant par rayons lumineux. On pourrait penser que l'espace-temps local de chaque observateur est donné, en première approximation, par l'espace tangent à une variété Lorentzienne au sens habituel. Mais comme un atome dans un champ de gravitation voit en général ses fréquences d'émission modifiées (Pound et Rebka), il faut conclure que, même s'il est au repos, son système propre n'est pas identique à celui du laboratoire, puisque ses unités en diffèrent. La donnée d'une variété de la Relativité Générale ne suffit donc pas pour l'interprétation physique de celle-ci. Même en l'absence d'accélérations, le système propre \emph{n'est pas en général un système d'inertie de la Relativité restreinte, mais un pseudo-système d'inertie au sens de la C-équivalence}. Nous résumons ce raisonnement par notre \textbf{second point}  :
\begin{quote}
Il faut distinguer variété des événements et variété des observateurs. Chaque observateur dispose d'une variété d'événements qu'il peut repérer. Il peut communiquer avec certains [p.~104] autres observateurs et alors, il existe des correspondances partielles entre les événements repérés par l'un et par l'autre. La variété des observateurs est l'ensemble de ces variétés.
\end{quote}

\section{L'observateur et sa variété des événements}

On pourrait penser que l'introduction d'un calcul d'opérateurs dispenserait  d'examiner les notions d'espace et de temps attachées à un observateur. Ainsi, von Neumann affirmait en 1933, alors même que la notion moderne de variété n'avait pas été complètement élaborée \cite{whitney}, que les mathématiques qu'il connaissait suffisaient pour représenter tous les apports de la nouvelle mécanique :
\begin{quote}
    \og En son temps, la mécanique de {Newton} a contribué à fonder sur des bases solides le calcul infinitésimal, lequel indubitablement n'était pas, à cette époque, exempt de contradictions internes. On pourrait être tenté de croire que, d'une façon analogue, la mécanique quantique suggère elle aussi une refonte complète de notre \og analyse à un nombre infini de variables \fg\ ; en d'autres termes on pourrait croire qu'en fin de compte il nous faille modifier l'instrument mathématique et non pas la théorie physique qui l'utilise. Or il n'en est rien ; nous démontrerons qu'on peut, avec tout autant de clarté et d'homogénéité qu'auparavant, bâtir la théorie des transformations sur des fondements mathématiques absolument rigoureux. Le développement correct que nous envisageons ne consistera nullement à expliciter et à préciser la méthode de   {Dirac}, mais nous obligera dès le début à utiliser une théorie entièrement différente, à savoir la théorie spectrale des opérateurs de {Hilbert}.\fg\   \cite[p. 2]{von-neumann}
\end{quote}
Mais la notion d'observateur est antérieure à tout calcul d'opérateurs, car ceux-ci représentent les résultats d'opérations effectuées par un observateur physique qui leur préexiste. C'est pourquoi nous ne pouvons suivre von Neumann. Ici encore, Louis de Broglie avait vu plus loin :
\begin{quote}
\og La fonction $\Psi$, en effet, ne représente pas quelque chose qui aurait son siège en un point de l'espace à un instant donné : elle représente, prise dans son ensemble, l'état des connaissances d'un observateur à l'instant considéré sur la réalité physique qu'il étudie : rien d'étonnant alors à ce que la fonction $\Psi$ varie d'un observateur à un autre, puisqu'ils [p.~105] ne possèdent pas en général les mêmes renseignements sur le monde qui les entoure, n'ayant point effectué les mêmes observations, ni les mêmes mesures.\fg\ \cite[p.~150]{p-m}
\end{quote}
Nous prolongeons ce point en suggérant que la conception de l'onde de matière conduit à étendre la notion d'observateur à des systèmes distribués d'oscillateurs physiques, sans référence à un sujet conscient.

Louis de Broglie avait souligné une seconde incohérence de l'ancienne mécanique : elle ne permet pas de penser le mouvement :
\begin{quote}
\og Le cadre de l'espace et du temps est essentiellement statique : un corps, une entité physique, qui a une localisation exacte dans l'espace et le temps est, par le fait même, privé de toute propriété évolutive ; au contraire un corps qui évolue, qui est doué de propriétés dynamiques, ne peut être véritablement rattaché à aucun point de l'espace et du temps. Ce sont des remarques philosophiques qui remontent à Zénon d'\'Elée et à ses disciples.  \cite[p.~138--139]{p-m} [\dots] la Mécanique classique [\dots] ne sait représenter le mouvement que par des positions successives sur une courbe continue [contrairement à] la Mécanique ondulatoire, qui, elle, sait représenter la mobilité sans aucune préoccupation de localisation par l'image analytique de l'onde place monochromatique.\fg \cite[p.~202--203]{p-m}
\end{quote}
Nous proposons en conséquence que le rapport logique entre l'espace-temps et l'onde monochromatique soit inversé : l'espace-temps n'est plus donné \emph{a priori}, c'est chaque onde de matière qui détermine un espace-temps qui lui est propre. Il faut alors lui associer une carte locale des événements mesurables par coïncidence avec certains des oscillateurs qui constituent l'onde. La multiplicité de ces cartes est irréductible, sauf à postuler un observateur unique qui aurait accès à tous les événements de tous les systèmes. Aucun observateur physique ne semble avoir cette propriété. Le changement de l'état de mouvement impose une modification du système propre de sorte que le mouvement n'est plus représenté comme une sucession de positions dans un espace ambiant.

Lorsque deux observateurs inertiels sont en translation uniforme l'un par rapport à l'autre, la Relativité restreinte montre comment on peut établir une correspondance entre leurs systèmes propres, qui autorise, par abus de langage, à les considérer comme plongés dans le même espace de Minkowski. Lorsque des observateurs éloignés, quelconques, ne [p.~106] communiquent que par rayons lumineux, Kichenassamy a donné un protocole dans le cadre de la C-équivalence \cite{K-ihp-ceq} qui leur permet d'établir une correspondance partielle entre leurs observations. Bien qu'il ait précisé alors que la C-équivalence permettait de \og compléter la théorie de J. L. Synge du décalage vers le rouge et de distinguer dans l'effet global l'effet gravitationnel de l'effet Doppler \fg \cite[(i), résumé]{c-equiv}, le point de vue diffère profondément de celui de Synge, qui considérait au contraire l'espace-temps comme un nouvel absolu \cite{synge-equiv}.

Si les différents observateurs ne peuvent être inclus dans un unique espace ponctuel donné \emph{a priori}, la notion usuelle de variété comme espace de points perd sa pertinence. Examinons sous cet angle l'axiomatique usuelle des variétés \cite{k-n}.
 Dans celle-ci, on se donne (i) un espace topologique $M$ supposé séparé au sens de Hausdorff (ce sera l'ensemble des \og points \fg\ de la variété) ; (ii) des ouverts $U_i$ dont la réunion est $M$ ; (iii) des homéomorphismes $\varphi_i : U_i\to V_i\subset\mathbb R^4$ (chacun des $(U_i,\varphi_i)$ constituant une \og carte locale \fg), tels que (iv) les applications $\varphi_j\circ\varphi_i^{-1}$ soient de classe $C^\infty$ dans leurs domaines de définition (ce sont les \og changements de carte\fg).

L'axiome (i) est proprement invérifiable, sauf à avoir déjà construit $M$ par d'autres moyens. Une modification minimale du formalisme habituel consisterait à considérer que la variété des observateurs est elle-même un ensemble de \emph{variétés lorentziennes au sens usuel} $\mathcal{V}_i$ homéomorphes via des applications $\varphi_i$ à des ouverts $V_i$ de $\mathbb{R}^4$. Chacune des $\mathcal{V}_i$ représente la géométrie locale d'un observateur $\mathcal O_i$. On supposerait en outre que \emph{certaines} des applications $\varphi_j\circ\varphi_i^{-1}$ soient bien définies sur un domaine non vide, et de classe $C^\infty$. Les points des $\mathcal{V}_i$ représenteraient les événements effectivement repérables par l'observateur $\mathcal O_i$. L'existence de tels difféomorphismes de passage entre observateurs devra refléter la possibilité effective pour ceux-ci de coordonner leurs observations. Il n'y a aucun moyen physique de se placer dans un système qui pourrait avoir accès à tous les points de vue de tous les observateurs à la fois.

Il conviendrait donc de déterminer les caractéristiques de ces systèmes locaux associés aux différents états de mouvement de systèmes accélérés, et particulièrement leur extension spatiale et leurs fréquences caractéristiques. Examinons quelques situations favorables à une telle étude. [p.~107]

\section{Application à quelques situations concrètes}

Un aspect du problème -- l'extension spatiale du système propre --, semble avoir été identifié dès 1957. En effet, Lennuier avait montré dans sa thèse, sur le cas de la vapeur de mercure, que la résonance optique pouvait être provoquée par une lumière de fréquence légèrement différente de celle de la résonance, et que l'émission s'effectuait après un léger retard. C'est pour expliquer la fraction d'émission incohérente observée par Lennuier que Kichenassamy (1951) avait suggéré que l'atome excité serait, bien que l'hamiltonien relativiste fût linéaire, le lieu de transitions mettant en jeu deux quanta \`a la fois \cite[pp.~863--864]{mq}, et en avait estimé les durées de vie par une modification originale de la méthode de Wigner-Weisskopf (ibid.). Peu après, il avait proposé que la même analyse pourrait s'appliquer aux collisions entre atomes et électrons : ils formeraient ensemble un \og complexe généralement virtuel au sens de la Mécanique Ondulatoire \fg\ \cite[p.~1035]{coll}. Or, il avait justifié son approche en remarquant que, \og s'il est permis en diffusion optique de supposer que les dimensions du système atomique en interaction sont très petites par rapport à la longueur d'onde incidente de façon que les variations de la phase dans l'expression du potentiel vecteur soient négligeables à l'intérieur de l'atome, il ne l'est pas entièrement, et que précisément les processus ``simultanés'' ont la prétention de tenir compte des déphasages éventuels \fg. L'extension spatiale non négligeable des processus en cause dans la résonance optique sera confirmée en 1957, lorsque Guiochon, Blamont et Brossel notèrent que certains des résultats antérieurs dépendaient \og de façon critique de la géométrie du montage\fg, et particulièrement du \og volume et [de la] forme de la cellule \fg\ \cite[p.~99]{guiochon}. Ils en déduisirent que l'interaction s'effectuait sur des distances relativement importantes [\dots] (de l'ordre du millimètre) \fg\ \cite[p.~100]{guiochon}. Ces observations sont cohérentes avec l'hypothèse selon laquelle l'atome définirait un système étendu d'oscillateurs. Il serait souhaitable de répéter ces expériences.

Considérons maintenant trois exemples de systèmes accélérés de dimensions appréciables. Le plus simple, et le plus proche des idées de Louis de Broglie, est sans doute fourni par les interactions entre électron et molécule que l'on a récemment interprétées comme des \og interférences à un seul électron \fg\  \cite{fremont}. Il est manifeste que le système propre de la molécule libre diffère de celui du complexe qu'elle forme avec l'électron, et que la transition de l'un à l'autre ne peut s'exprimer en termes de mouvements rectilignes et uniformes. Il est donc possible que les fréquences d'émission de la molécule qui, au repos, ne seraient pas affectées par l'addition de cet électron, soient modifiées par l'accélération subie pendant la capture ou l'éjection de celui-ci. Malgré la relative faiblesse des vitesses, le temps de collision pourrait être suffisamment court pour engendrer une accélération importante, non constante de surcroît.

On trouve également des accélérations importantes dans le cadre du refroidissement Doppler. L'atome y subit plusieurs millions de cycles excitation/émission qui conduisent à une décélération moyenne pouvant atteindre $10^5 g$ sur une distance de l'ordre du mètre \cite[p.~709, col.~2]{cct-rmp}. La faible température assure une largeur Doppler faible. Cependant, dans ce cas, il s'agit sans doute d'un très grand nombre de brusques décélérations de très courte durée, donc encore plus importantes mais, pour cette raison même, plus difficiles à étudier. It n'est cependant pas exclu qu'une irradiation transverse d'un jet d'atomes pendant sa phase de décélération puisse permettre de déceler une modification de certaines fréquences d'émission, si le nombre d'événements était suffisamment important.

Enfin, l'analyse de l'électron accéléré \cite{K-cr-66a} a montré que le bilan d'énergie était modifié par le changement d'unités du système propre. En outre, toutes les fois que l'accélération n'est pas uniforme, on dispose d'une relation mathématique entre accélération mesurée dans le système du laboratoire et le temps propre au sens des formules de Frenet quadridimensionnelles \cite{K-cr-65d}. Ainsi, la mesure du décours de l'accélération détermine le temps propre d'un système non uniformément accéléré, mais uniformément suraccéléré. Il s'agit là d'une nouvelle méthode de mesure du temps propre à partir de la mesure de la trajectoire dans le système du laboratoire.

\section{Conclusion}

Le rapprochement de la Mécanique ondulatoire et de la C-équivalence conduit à postuler que le système propre d'un atome, d'un électron ou d'une molécule, éventuellement en interaction avec un champ, est un objet physique, définissant ses propres unités d'espace et de temps qui, même lorsqu'il est au repos dans le laboratoire, ne sont pas nécessairement celles des instruments \og standards \fg\ d'un système d'inertie de la Relativité restreinte. Il s'ensuit une modification de la notion de variété, qui ne se réduit à la notion usuelle que lorsque l'on suppose que l'espace ponctuel sous-jacent aux changements de cartes est commun à tous les observateurs. [p.~109]

Lorsque l'accélération n'est pas uniforme, il existe, même en Relativité restreinte, une relation entre temps propre et accélération, et donc entre temps propre et temps du laboratoire, à condition de préciser la manière dont deux observateurs peuvent coordonner leurs mesures. On peut également envisager de mettre en évidence les caractéristiques physiques d'un tel C-système propre en le faisant interagir avec un rayonnement pendant ses phases d'accélération non uniforme.
Nous avons indiqué plusieurs situations dans lesquelles les caractérisiques physiques de ces systèmes propres d'un nouveau type sont accessibles à l'expérience.

La C-équivalence semble être la seule théorie qui n'assimile pas le mouvement à un déplacement géométrique dans un espace indépendant de l'observateur, et qui permette de tirer des conséquences physiques et mathématiques de cette distinction. On a montré que son extension à la microphysique ouvrait la voie à une expression mathématique de la vision de Louis de Broglie.


\begin{thebibliography}{99}

\bibitem{these-LdB} Louis de Broglie, \emph{Recherches sur la Théorie des Quanta : Réédition du texte de 1924}, Masson, Paris, 1963.

\bibitem{K-ihp-ceq}
    S.~Kichenassamy,
    \newblock Compl\'ements \`a l'interpre\'etation physique de la {R}elativit\'e
    g\'en\'erale: Applications.
    \newblock {\em Annales de l'Inst.\ H. Poincar\'e, Section A (N. S.)}, {\bf 1} (2), 129--145, (1964). Voir également {\em C. R. Acad. Sci. Paris}, {\bf 258}, 470--473, (1964).

\bibitem{c-equiv} La C-équivalence est l'aboutissement d'un examen serré de toutes les théories antérieures, qui n'a pas entièrement été publié. On trouvera les principaux arguments dans les articles cités note \cite{K-ihp-ceq} et dans les travaux suivants : (i) {\em Annales de l'Inst.\ H. Poincar\'e, Section A (N. S.)}, \textbf{4} (2), 139--158, (1966) ; (ii) \og La {R}elativit\'e {G}\'en\'erale \fg, {\em Relativit\'e et quanta : les grandes th\'eories de la Physique moderne}, Masson, Paris, 1968, chapitre~II, pages 27--58 ; (iii) \og Remarks on the present state of {General Relativity} \fg, {\em Symposia on Theoretical Physics}, A.~Ramrakrishnan (ed.), vol.~5, Plenum Press, New York, 1967, p.~107--137 ; (iv) {\em Fluides et champ gravitationnel en Relativit\'e g\'en\'erale}, \'Editions du CNRS, Paris, 1969, pp.~241-245. Voir également, pour une introduction sommaire, Satyanad Kichenassamy, \emph{Annales de la Fondation Louis de Broglie}, {\bf 46}, 131--151, (2016). L'hommage de la Société Asiatique de Paris, annoncé dans ce travail, est paru dans le {\em Journal Asiatique}, {\bf 306} (1), 85--99, (2018).

\bibitem{cct-laloe}
    Claude Cohen-Tannoudji, dans \emph{Comprenons-nous vraiment la M\'ecanique Quantique ?}, F. Lalo\"e, EDP Sciences/CNRS \'Editions, Paris, 2011, p.~xi.

\bibitem{temps-nietzsche} Cette assertion ne va pas de soi : Pour Nietzsche : \og Le temps en soi est une absurdité, il n'a de temps que pour un être doué de sensation. De même pour l'espace.\fg\ (Zeit an sich ist ein Unsinn : nur für ein empfindenden [p.~110] Wesen gibt es Zeit. Ebenso Raum. \emph{Das Philosophen\-buch}, Aubier, 1969, I, 121, p. 118. Trad. A. K. Marietti, légèrement modifiée). Mais c'est là confondre mesure et sensation. La Relativité confirme que \og le temps en soi est une absurdité \fg, mais par un argument entièrement différent.

\bibitem{sk-cours} Ce point a été développé très clairement dans les cours de S. Kichenassamy à Paris dans les années 1960-70, malheureusement inédits. On trouvera un aper\c cu dans \cite[(ii) et (iii)]{c-equiv} et dans \emph{Annales de la Fond. L. de Broglie}, \textbf{5} (3), 1980, 191--202. Les éléments essentiels de notre travail s'y trouvent déjà.

\bibitem{anderson} J. L. Anderson, Revista Mexicana de Fisica, \textbf{III} (3), 176--184, (1957), E. Wigner, 
Rev.\ Mod.\ Phys. \textbf{29}, 255--268, (1957). On trouvera une synthèse des principaux arguments dans \cite[(iii)]{c-equiv} .

\bibitem{whitney} On fait généralement remonter la notion moderne de variété différentielle à H. Whitney, \og Differentiable manifolds \fg, \emph{Ann. Math.}, \textbf{37} (3), 645-680, (1936). Il montre qu'une version simplifiée de la notion de variété selon Cartan équivaut à la notion de sous-variété d'un espace $\mathbb{R}^n$ de grande dimension. Or, cet espace ambiant n'a aucune signification physique. Cette version simplifiée ne convient donc pas.

\bibitem{cartan-varietes} \emph{Le\c cons sur la géométrie des espaces de Riemann}, Gauthier-Villars, Paris, 2\ieme\ édition, revue et augmentée, 1946 (nouveau tirage : 1951), p. 57. Encore aujourd'hui, les variétés ne sont traitées dans Bourbaki que dans un fascicule de résultats: N. Bourbaki, \emph{Variétés différentielles et analytiques}, (fascicule de résultats, deuxième édition corrigée), Hermann, Paris, 1971. C'est à notre connaissance le seul traité qui envisage que le modèle local puisse dépendre du point : les cartes locales sont de la forme $(X, \varphi, E)$ où $\varphi$ est une bijection sur un ouvert de $E$, où ce dernier \emph{dépend} de la carte (\S\ 5.1.1., p. 34). Cartan allait plus loin, en envisageant des voisinages \og infinitésimaux\fg\ d'un point, notion sans équivalent moderne. Nous ajoutons à cette exigence une autre: les systèmes attachés à chaque observateur doivent être construits par un protocole physique.

\bibitem{cartan-aens} Cartan, \'Elie,  
    \emph{Annales Scientifiques de l'E.N.S.}, \textbf{40}, (1923), 325-412 ; \textbf{41}, (1923), 1-25 ; \textbf{42}, (1923), 17-88. Faut-il rappeler qu'\'Elie Cartan était l'un des membres du jury de thèse de Louis de Broglie ?


\bibitem{mq} Kichenassamy, S.,
\og Interaction entre matière et rayonnement au voisinage de la résonance
optique\fg,\
{\em  Journal de Physique et Le Radium}, {\bf 12}, 863--867, (1951).

\bibitem{coll} S. Kichenassamy,
{\em C. R. Acad. Sci. Paris},
{\bf 234}, 1035-1036 et 1530--1532, (1952).






\bibitem{guiochon}
Marie-Anne Guiochon, Jacques E. Blamont, Jean Brossel,
{\em J. Phys. Radium}, {\bf 18}, 99--108 (1957).

\bibitem{fremont}
Fran\c cois Fr\'emont,
\emph{Young-type Interferences with Electrons}, Springer, 2014.

\bibitem{cct-rmp}
Claude Cohen-Tannoudji,
{\em Rev. Mod. Phys.}, {\bf 70} (3), 707--719, (1998).

\bibitem{K-cr-66a}
    S.~Kichenassamy,
    \newblock {\em C. R. Acad. Sci. Paris},
    \textbf{B262}, 1591--1593, (1966).

\bibitem{K-cr-65d}
    S.~Kichenassamy,
    {\em C. R. Acad. Sci., Paris}, {\bf 260}, 3865--3868, (1965).


\bibitem{von-neumann}
J. von Neumann, \emph{Mathematische Grundlagen der Quantenmechanik}, Springer, 1932. Nous citons la traduction d'A. Proca  (Alcan, Paris, 1946. Réimpr. : Gabay, Sceaux, 1988).
[p.~111]
\bibitem{p-m}
Louis de Broglie, \emph{Physique et microphysique}, Albin Michel, Paris, 1947.

\bibitem{synge-equiv} \og The Principle of Equivalence performed the essential office of midwife at the birth of general relativity, [...] I suggest that the midwife be now buried with appropriate honors and the facts of absolute space-time faced.\fg\ J. L. Synge, \emph{ Relativity : The General Theory}, North-Holland, Amsterdam, 1960, p., ix-x.

\bibitem{k-n}
    Shoshichi Kobayashi et Katsumi Nomizu,
    \emph{Foundations of Differential Geometry, I},
    Wiley-Interscience, New York, 1963.

\end{thebibliography}
\end{document}